\documentstyle[12pt,epsfig]{article}
\newcommand{\be}{\begin{equation}}
\newcommand{\ee}{\end{equation}}
\newcommand{\bea}{\begin{eqnarray}}
\newcommand{\eea}{\end{eqnarray}}

\newcommand{\bfk}{\mbox{\boldmath $k$}}
\newcommand{\bfq}{\mbox{\boldmath $q$}}
\newcommand{\pup}{p^\uparrow}
\newcommand{\pdown}{p^\downarrow}
\newcommand{\qup}{q^\uparrow}
\newcommand{\qdown}{q^\downarrow}

\newcommand{\bfp}{\mbox{\boldmath $p$}}
\newcommand{\bfP}{\mbox{\boldmath $P$}} 
\newcommand{\Lup}{\Lambda^\uparrow} 
 
\newcommand{\Aup}{A^\uparrow} 
\newcommand{\hup}{h^\uparrow} 
\newcommand{\hdown}{h^\downarrow} 
\newcommand{\nd}{\noindent}
\def\lsim{\mathrel{\rlap{\lower4pt\hbox{\hskip1pt$\sim$}}\raise1pt\hbox{$<$}}}
\def\gsim{\mathrel{\rlap{\lower4pt\hbox{\hskip1pt$\sim$}}\raise1pt\hbox{$>$}}}
\newcommand{\NP}[1]{{\it Nucl.\ Phys.}\ {\bf #1}}
\newcommand{\ZP}[1]{{\it Z.\ Phys.}\ {\bf #1}}
\newcommand{\PL}[1]{{\it Phys.\ Lett.}\ {\bf #1}}
\newcommand{\PR}[1]{{\it Phys.\ Rev.}\ {\bf #1}}
\newcommand{\PRL}[1]{{\it Phys.\ Rev.\ Lett.}\ {\bf #1}}

\newcommand{\IJMP}[1]{{\it Int.\ J.\ Mod.\ Phys.}\ {\bf #1}}

\begin{document}
%%%%%%%%%%%%%%%%%%%%%%%%%%%%%%%%%%%%%%%%%%%%%%%%%%%%%%%%%%%%%%%%%%%%%%%%%%%%%%
\begin{flushright} 
%DFTT 28/2001 \\ 
INFNCA-TH0208 \\ 
%hep-ph/0210371 \\ 
\end{flushright} 
\vskip 1.5cm
\begin{center}
{\bf Transverse single spin asymmetries in Drell-Yan processes}\\
\vskip 0.8cm
{\sf M.~Anselmino$^1$, U.~D'Alesio$^2$, F.~Murgia$^2$}
\vskip 0.5cm
{\it $^1$ Dipartimento di Fisica Teorica, Universit\`a di Torino and \\
          INFN, Sezione di Torino, Via P. Giuria 1, I-10125 Torino, Italy}\\
\vspace{0.3cm}
{\it $^2$ INFN, Sezione di Cagliari and Dipartimento di Fisica,  
Universit\`a di Cagliari,\\
C.P. 170, I-09042 Monserrato (CA), Italy} \\
\end{center}

\vspace{1.5cm}

\begin{abstract}
Recently, it has been shown, contrary to previous beliefs, that 
the $\bfk_\perp$ distribution of quarks in a transversely polarized proton 
can be asymmetric. This ``Sivers effect'' had already been used to explain 
transverse single spin asymmetries (SSA) observed in inclusive pion 
production, $\pup \, p \to \pi \, X$ and $\bar{p}^\uparrow \, p \to \pi \, X$. 
In such channels, however, other mechanisms, like the ``Collins effect''
(a $\bfk_\perp$ asymmetric fragmentation of a transversely polarized quark 
into pions), may generate SSA. The Sivers asymmetry is used here 
to compute SSA in Drell-Yan processes; in this case, by considering the 
differential cross-section in the lepton-pair invariant mass, rapidity and 
transverse momentum, other mechanisms which may originate SSA cannot 
contribute. Estimates for RHIC experiments are given.
\end{abstract}

\vspace{0.6cm}

%{}~~~PACS numbers: %13.88.+e, 13.60.-r, 13.15.+g, 13.85.Ni
%%%%%%%%%%%%%%%%%%%%%%%%%%%%%%%%%%%%%%%%%%%%%%%%%%%%%%%%%%%%%%%%%%%%%%%%%%%%%% 
\newpage 
%%%%%%%%%%%%%%%%%%%%%%%%%%%%%%%%%%%%%%%%%%%%%%%%%%%%%%%%%%%%%%%%%%%%%%%%%%%%%% 
\pagestyle{plain} 
\setcounter{page}{1} 
\nd 
{\bf 1. Introduction} 
\vskip 6pt 
Single spin asymmetries in high energy inclusive processes are a unique
testing ground for QCD; they cannot originate from the simple spin pQCD 
dynamics -- dominated by helicity conservation -- but need some non
perturbative chiral-symmetry breaking in the large distance physics. 
Experiments are rich of puzzling single spin results and more are expected 
to come; new theoretical ideas and models have just appeared in the 
literature and a good understanding seems to emerge.    

Among the best known transverse single spin asymmetries (SSA) let us 
mention:

\begin{itemize}

\item
the old problem of the large polarization of $\Lambda$'s and other 
hyperons produced in the scattering of unpolarized nucleons,
$p \, N \to \Lup \, X$ \cite{lam};

\item
the large asymmetry 
\be
A_N = \frac{d\sigma^\uparrow - d\sigma^\downarrow}
           {d\sigma^\uparrow + d\sigma^\downarrow} \label{asy}
\ee
observed for pion inclusive production, in $\pup \, p \to \pi \, X$ and  
$\bar p^\uparrow \, p \to \pi \, X$ processes \cite{e704};

\item
the similar azimuthal asymmetry observed in semi-inclusive DIS processes, 
$\ell \, \pup \to \ell \, \pi \, X$ \cite{herm}.

\end{itemize}

All of these cannot be related to the elementary dynamics, but rather
to non perturbative aspects of the nucleon and hadron structures. 

Among the several attempts \cite{prag} to explain the data within QCD 
dynamics we focus here on a phenomenological approach based on the 
generalization of the factorization theorem with the inclusion of parton 
intrinsic motion $\bfk_\perp$ inside a nucleon and of hadrons relatively 
to the fragmenting parton. The cross-section for a generic process 
$A\,B \to C\,X$ then reads:
\be
d\sigma = \sum_{a,b,c} \hat f_{a/A}(x_a,\bfk_{\perp a}) \otimes 
\hat f_{b/B}(x_b, \bfk_{\perp b}) \otimes
d\hat\sigma^{ab \to c \dots}(x_a, x_b, \bfk_{\perp a}, \bfk_{\perp b}) 
\otimes \hat D_{C/c}(z, \bfk_{\perp C})
\label{ltgen}
\ee
where the $\hat f$'s are the $\bfk_\perp$ dependent parton distributions 
and the $\hat D$'s the $\bfk_\perp$ dependent fragmentation functions.
  
The above QCD factorization theorem -- with unintegrated $\bfk_\perp$ 
dependent distribution and fragmentation functions -- has never been 
formally proven in general \cite{col}, but only for the Drell-Yan process
(which is the issue of this paper) and for the two-particle inclusive 
cross-section in $e^+e^-$ annihilation (somehow a time-reversed Drell-Yan
process) \cite{css}. However, Eq. (\ref{ltgen}) has been widely used in 
the literature, starting from the pioneering work of Feynman, Field and
Fox \cite{fff}. Actually, several papers have recently shown that
intrinsic $\bfk_\perp$'s are indeed necessary in order to be able to 
explain, within pQCD and the factorization scheme, data on large $p_T$
production of pions and photons \cite{pkt}; without them the theoretical
(collinear) computations would give results in some cases much smaller 
(up to a factor 100) than experiments.     

When dealing with polarized processes the introduction of $\bfk_\perp$ 
dependences opens up the way to many possible spin effects; these can be 
summarized, at leading twist, by new polarized distribution functions,
\bea
\Delta^Nf_{q/\pup}  \!\!\! &\equiv& \!\!\!
\hat f_{q/\pup}(x, \bfk_{\perp})-\hat f_{q/\pdown}(x, \bfk_{\perp}) =  
\hat f_{q/\pup}(x, \bfk_{\perp})-\hat f_{q/\pup}(x, - \bfk_{\perp}) 
\label{delf1} \\
\Delta^Nf_{\qup/p}  \!\!\! &\equiv& \!\!\!
\hat f_{\qup/p}(x, \bfk_{\perp})-\hat f_{\qdown/p}(x, \bfk_{\perp}) =
\hat f_{\qup/p}(x, \bfk_{\perp})-\hat f_{\qup/p}(x, - \bfk_{\perp})  
\label{delf2}
\eea
and new polarized fragmentation functions,
\bea 
\!\!\!\! \Delta^N D_{h/\qup} \!\!\!\! &\equiv& \!\!\!\!
\hat D_{h/\qup}(z, \bfk_{\perp}) - \hat D_{h/\qdown}(z, \bfk_{\perp}) =
\hat D_{h/\qup}(z, \bfk_{\perp})-\hat D_{h/\qup}(z, - \bfk_{\perp}) 
\label{deld1} \\
\!\!\!\! \Delta^N D_{\hup/q} \!\!\!\! &\equiv& \!\!\!\!
\hat D_{\hup/q}(z, \bfk_{\perp}) - \hat D_{\hdown/q}(z, \bfk_{\perp}) =
\hat D_{\hup/q}(z, \bfk_{\perp})-\hat D_{\hup/q}(z, - \bfk_{\perp}) 
\label{deld2} 
\eea
which have a clear meaning if one pays attention to the arrows denoting the
polarized particles. All the above functions vanish when $k_\perp=0$ and 
are na\"{\i}vely $T$-odd. The ones in Eqs. (\ref{delf2}) and (\ref{deld1}), 
when written in the helicity basis, relate quarks of different helicities 
and are chiral-odd, while the other two are chiral-even.   

Similar functions have been introduced in the literature with different 
notations: in particular there is a direct correspondence \cite{bm} between 
the above functions and the ones denoted, respectively, by: $f_{1T}^\perp$
\cite{dp}, $h_1^\perp$ \cite{dan}, $H_1^\perp$ and 
$D_{1T}^\perp$ \cite{dp,jac}.

In the recent comprehensive review paper on transverse quark polarization
\cite{bdr} the following notations are used: $\Delta^Nf_{q/\pup} \equiv
\Delta^T_0f$, $\Delta^Nf_{\qup/p} \equiv \Delta_T^0f$ and 
$\Delta^N D_{h/\qup} \equiv 2\Delta_T^0D_{h/q}$.

The fragmentation in Eq. (\ref{deld1}) is the Collins function \cite{col},
while the distribution in Eq. (\ref{delf1}) was first introduced by Sivers 
\cite{siv}.

Some of the above functions have been used for a phenomenological 
description of the observed SSA. Both Sivers \cite{nois} and Collins 
\cite{noic, ee} functions can explain the E704 data on $A_N$ \cite{e704}:
the Collins asymmetry might have been indirectly observed in HERMES data 
\cite{herm} and the polarizing fragmentation functions (\ref{deld2}) can 
describe \cite{noil} the $\Lambda$ polarization data \cite{lam}.  

Despite its successful phenomenology, the Sivers function was always
a matter of discussions and its very existence rather controversial; in fact
in Ref. \cite{col} a proof of its vanishing was given, based on time-reversal 
invariance. Ways out based on initial state interactions \cite{nois} or 
non standard time-reversal properties \cite{dra} were discussed, but
the situation remained uncertain and further phenomenology was
performed making use of the Collins function only \cite{noic}. A similar
criticism applied to the function in Eq. (\ref{delf2}).

Very recently a series of papers have resurrected Sivers asymmetry 
in its full rights. First, in Ref. \cite{bro1}, a quark-diquark model
calculation including final state effects has given an explanation of 
the HERMES azimuthal asymmetry different from the one originating 
from Collins function. As a consequence, Collins recognised \cite{col2}
that such a new mechanism is compatible with factorization and is due  
to the Sivers asymmetry (\ref{delf1}); re-examining his original proof 
of the vanishing of $\Delta^Nf_{q/\pup}$ he finds it to be invalidated
by the path-ordered exponential of the gluon field in the operator 
definition of parton densities.
The authors of Ref. \cite{ji} have confirmed the model results of 
\cite{bro1}; a second paper by Brodsky, Hwang and Schmidt \cite{bro2},
along the same lines of the first one, shows that initial state interactions 
give rise to SSA in Drell-Yan processes.

Some issues concerning factorizability and universality of these
effects are still open to debate; however, we feel now confident to use 
Sivers effects -- and equally all functions in 
Eqs. (\ref{delf1})-(\ref{deld2}) -- in SSA phenomenology. The natural 
process to test the Sivers asymmetry is Drell-Yan: in such a case there 
cannot be any effect in fragmentation processes and, by suitably integrating 
over some final configurations, also possible effects from transversely 
polarized quarks in an unpolarized proton, Eq. (\ref{delf2}), do not 
contribute. SSA in Drell-Yan processes are particularly important now, as 
ongoing or imminent experiments at RHIC will be able to measure them. 

The paper is organized as follows: in Section 2 we present the formalism 
necessary to discuss single spin asymmetries in Drell-Yan processes, 
taking into account the transverse motion of partons in nucleons.
Some simple analytical formulae are also derived, assuming simplified
functional $k_\perp$ dependences.  
In Section 3 we give some numerical estimates for SSA at RHIC, while
comments and conclusions are gathered in Section 4.

\vspace{18pt}
\goodbreak
\nd
{\bf 2. SSA in Drell-Yan processes, formalism}
\nobreak
\vspace{6pt}
\nobreak

Let us consider a Drell-Yan process, that is the production of 
$\ell^+\ell^-$ pairs in the collision of two hadrons $A$ and $B$; 
there is no need of any fragmentation function and Eq. (\ref{ltgen}) reads:
\be
d\sigma = \sum_{ab} \int \left[ dx_a \, d^2\bfk_{\perp a} 
\, dx_b \, d^2\bfk_{\perp b} \right] \, \hat f_{a/A}(x_a,\bfk_{\perp a}) \,
\hat f_{b/B}(x_b,\bfk_{\perp b}) \,
d\hat\sigma^{ab \to \ell^+\ell^-}
%(dx_a, d^2\bfk_{\perp a}, dx_b, d^2\bfk_{\perp b})
\label{dy1}
\ee
and, from the Sivers asymmetry of Eq. (\ref{delf1}), the difference between 
the single transverse spin dependent cross-sections $d\sigma^\uparrow$ for 
$A^\uparrow \, B \to \ell^+ \, \ell^- \, X$ and $d\sigma^\downarrow$ for
$A^\downarrow \, B \to \ell^+ \, \ell^- \, X$ is
\be
d\sigma^\uparrow - d\sigma^\downarrow = 
\sum_{ab} \int \left[ dx_a \, d^2\bfk_{\perp a} 
\, dx_b \, d^2\bfk_{\perp b} \right] \, 
\Delta^Nf_{a/A^\uparrow}(x_a,\bfk_{\perp a}) \,
\hat f_{b/B}(x_b,\bfk_{\perp b}) \,
d\hat\sigma^{ab \to \ell^+\ell^-}
%(dx_a, d^2\bfk_{\perp a}, dx_b, d^2\bfk_{\perp b}) 
\!.
\label{ddy1}
\ee

The elementary cross-section $d\hat\sigma$ for the process 
$a(p_a) \, b(p_b) \to \ell^+(p_+) \, \ell^-(p_-)$ is given by:
\be
d\hat\sigma = \frac{1}{2\hat s} \> \frac{d^3p_+}{2E_+} \> \frac{d^3p_-}{2E_-}
\> \frac{1}{(2\pi)^2} \> \delta^4(p_a + p_b - p_+ - p_-) \>
\overline{\left\vert \, M_{ab \to \ell^+\ell^-} \, \right\vert^2} \>. 
\label{ecs}
\ee     

We consider the differential cross-section in the variables
\be
\hat s \equiv M^2 = (p_a + p_b)^2 \equiv q^2 \quad\quad\quad
y = \frac 12 \ln \frac{q_0 + q_L}{q_0 - q_L} \quad \quad\quad \bfq_T \>,
\label{var}
\ee
that is the squared invariant mass, the rapidity and the transverse momentum 
of the lepton pair; $q_0$, $\bfq_T$ and $q_L$ are respectively the energy, 
transverse and longitudinal components, in the $A$-$B$ center of mass frame, 
of the four-vector $q = p_a + p_b = p_+ + p_-$.   

Using the relations:
\be
\frac{d^3p_-}{2E_-} = d^4p_- \, \delta(p_-^2) \quad\quad\quad
p_- = q - p_+ \quad\quad\quad
dM^2 \, dy = 2 \, dq_0 \, dq_L \>, 
\label{jac}
\ee
Eq. (\ref{ddy1}) can be written as
\bea
&&\frac{d^4\sigma^\uparrow}{dy \, dM^2 \, d^2\bfq_T} - 
\frac{d^4\sigma^\downarrow}{dy \, dM^2 \, d^2\bfq_T} = 
\frac{1}{2} \sum_{ab} \int \left[ dx_a \, d^2\bfk_{\perp a} 
\, dx_b \, d^2\bfk_{\perp b} \right] \nonumber \\
&&\Delta^Nf_{a/A^\uparrow}(x_a,\bfk_{\perp a}) \,
\hat f_{b/B}(x_b,\bfk_{\perp b}) \, \delta^4(p_a + p_b - q) 
\, \hat\sigma^{ab}_0 \label{ddy2} 
\eea
where $\hat\sigma^{ab}_0$ is the total cross-section for the 
$ab \to \ell^+\ell^-$ process:
\be
\hat\sigma_0^{ab} = \int \frac{d^3p_+}{2E_+} \> \frac{1}{(2\pi)^2}
\frac{1}{2M^2} \> \delta((q-p_+)^2) \> 
\overline{\left\vert \, M_{ab \to \ell^+\ell^-} (p_+,q) \, \right\vert^2} 
\>. \label{ecs0}
\ee

Analogously
\bea
&&\frac{d^4\sigma^\uparrow}{dy \, dM^2 \, d^2\bfq_T} + 
\frac{d^4\sigma^\downarrow}{dy \, dM^2 \, d^2\bfq_T} = 
\sum_{ab} \int \left[ dx_a \, d^2\bfk_{\perp a} 
\, dx_b \, d^2\bfk_{\perp b} \right] \nonumber \\
&& \hat f_{a/A}(x_a,\bfk_{\perp a}) \,
\hat f_{b/B}(x_b,\bfk_{\perp b}) \, \delta^4(p_a + p_b - q) 
\, \hat\sigma^{ab}_0 \label{udy2} 
\eea
which is twice the unpolarized cross-section.

For Drell-Yan processes the dominating electromagnetic elementary interaction 
is $q \bar q \to \gamma^* \to \ell^+\ell^-$, so that $a,b = q,\bar q$ with 
$q = u, \bar u, d, \bar d, s, \bar s$ and: 
\be
\hat\sigma^{q\bar q}_0 = \frac{4\,\pi \, \alpha^2 \, e_q^2}{9\,M^2} \>\cdot
\label{s0}
\ee
  
Eqs. (\ref{ddy2}) and (\ref{udy2}) allow to compute a single spin 
asymmetry $A_N$, Eq. (\ref{asy}), for Drell-Yan processes and for differential 
cross-sections measuring the lepton pair invariant mass $M$, rapidity $y$
and transverse momentum $\bfq_T$; notice that we do not look at the 
angular distribution of the lepton pair production plane, which is 
integrated over.  

Let us now fix in more details our kinematical configuration. We take the 
hadron $A$ as moving along the positive $z$-axis, in the $A$-$B$ c.m. frame 
%let us define the $A$-$\gamma^*$ plane as the $x$-$z$ plane and let us 
and measure the transverse polarization of hadron $A$, $\bfP_{\!A}$, along the 
$y$-axis, as shown in Fig. 1. Neglecting masses, the four-momenta of hadrons 
and partons are:
\be
p_A = \frac{\sqrt s}{2}\,(1, 0, 0, 1) \quad\quad\quad     
p_B = \frac{\sqrt s}{2}\,(1, 0, 0,-1)
\ee
\be
p_a = x_a \, \frac{\sqrt s}{2} \left( 1 + \frac{k_{\perp a}^2}{x_a^2 s}, \>
\frac{2\bfk_{\perp a}}{x_a \sqrt s}, \> 
1 - \frac{k_{\perp a}^2}{x_a^2 s} \right)
\ee
\be
p_b = x_b \, \frac{\sqrt s}{2} \left( 1 + \frac{k_{\perp b}^2}{x_b^2 s}, \>
\frac{2\bfk_{\perp b}}{x_b \sqrt s}, \>
-1 + \frac{k_{\perp b}^2}{x_b^2 s} \right)
\ee
with
\be
q = p_a + p_b = (q_0, \, \bfq_T, \, q_L) = \left( 
\sqrt {M^2 + q_T^2}\, \cosh y, \> \bfq_T, \>  
\sqrt {M^2 + q_T^2}\, \sinh y \right) \>,
\ee
where the lepton pair rapidity $y$ and invariant mass $M$ are defined
in Eq. (\ref{var}). 

%\begin{figure}[t]
%\begin{center}
%\epsfig{file=dysivers_1.eps,width=1.0\textwidth}
%\caption{\small{Our kinematical configuration. The $\gamma^*$ four-momentum
%defines all our observables; the dependence on the angle between the 
%$A$-$\gamma^*$ and the $\gamma^*$-$(\ell^+\ell^-)$ planes is integrated over 
%in Eq. (12) and (14).}}
%\end{center}
%\end{figure}

The four-momentum conservation $\delta$ function of 
Eqs. (\ref{ddy2}) and (\ref{udy2}) contains the factors:
\bea
&&\frac{1}{2} \, \delta(E_a + E_b - q_0) \, \delta(p_{za} + p_{zb} - q_L) 
 = \nonumber \\ 
&&\frac{1}{2} \, \delta \! \left( (x_a + x_b)\frac{\sqrt s}{2} + 
\left[ \frac{k_{\perp a}^2}{x_a s} + \frac{k_{\perp b}^2}{x_b s} \right]
\frac{\sqrt s}{2} - q_0 \right) \times \nonumber \\
&& \delta \! \left( (x_a - x_b)\frac{\sqrt s}{2} - 
\left[ \frac{k_{\perp a}^2}{x_a s} - \frac{k_{\perp b}^2}{x_b s} \right]
\frac{\sqrt s}{2} - q_L \right) \>. \label{ndel}
\eea 

We shall consider kinematical regions such that:
\be
q_T^2 \ll M^2 \quad\quad\quad k_{\perp a,b}^2 \simeq q_T^2 \>,
\ee
where Eq. (\ref{ndel}) simplifies into the usual collinear 
condition:
\be
\frac{1}{2} \, \delta(E_a + E_b - q_0) \, \delta(p_{za} + p_{zb} - q_L)
= \frac{1}{s} \, \delta \! \left( x_a - \frac{M}{\sqrt s} \, e^y \right) \, 
\delta \! \left( x_b - \frac{M}{\sqrt s} \, e^{-y} \right) \>. \label{odel}
\ee
In such a region $[q_T^2 \ll M^2, \> q_T \simeq k_\perp]$ then one has
\be
A_N = \frac
{\sum_q e_q^2 \int d^2\bfk_{\perp q} \, d^2\bfk_{\perp \bar q} \>
\delta^2(\bfk_{\perp q} + \bfk_{\perp \bar q} - \bfq_T) \>
\Delta^Nf_{q/\Aup}(x_q, \bfk_{\perp q}) \>
\hat f_{\bar q/B}(x_{\bar q}, \bfk_{\perp \bar q})}
{2 \sum_q e_q^2 \int d^2\bfk_{\perp q} \, d^2\bfk_{\perp \bar q} \>
\delta^2(\bfk_{\perp q} + \bfk_{\perp \bar q} - \bfq_T) \>
\hat f_{q/A}(x_q, \bfk_{\perp q}) \>
\hat f_{\bar q/B}(x_{\bar q}, \bfk_{\perp \bar q})} \label{ann}
\ee
with $x_q$ and $x_{\bar q}$ fixed by Eq. (\ref{odel}) with $a,b = q, \bar q$  
and $q = u, \bar u, d, \bar d, s, \bar s$.    

\vspace{12pt}
\goodbreak
\nd
{\bf 2.1 Other mechanisms for SSA in Drell-Yan processes}
\nobreak
\vspace{4pt}
\nobreak

Before discussing further Eq. (\ref{ann}), let us comment on other possible 
origins of SSA. Let us consider first the SSA generated by the distribution 
function in Eq. (\ref{delf2}), as compared with Sivers mechanism, 
Eq. (\ref{delf1}), which we are considering here. 
The main point is that, as we have seen, Eq. (\ref{delf1}) gives a 
contribution to $A_N$ of the type:
\be
\sum_q \Delta^N f_{q/A^{\uparrow}}(x_a,\bfk_{\perp a}) \otimes 
\hat f_{\bar q/B}(x_b, \bfk_{\perp b}) \otimes
d\hat\sigma^{q\bar q  \to \ell^+\ell^-}  \label{ansiv}
\ee
where $d\hat\sigma$ is the elementary unpolarized cross-section, 
while Eq. (\ref{delf2}) leads to a con\-tri\-bu\-tion of the kind \cite{dan}   
\be
\sum_q h_{1q}(x_a,\bfk_{\perp a}) \otimes 
\Delta^N f_{\bar q^\uparrow/B}(x_b, \bfk_{\perp b}) \otimes
d\Delta \hat\sigma^{q\bar q  \to \ell^+\ell^-}  \label{andan}
\ee
where $h_{1q}$ is the transversity of quark $q$ (inside hadron $A$) 
and $d\Delta \hat\sigma$ is the double transverse spin asymmetry 
$d\hat\sigma^{\uparrow\uparrow} - d\hat\sigma^{\uparrow\downarrow}$.
Such an elementary asymmetry has a $\cos2\phi$ dependence \cite{dan,rhic}, 
where $\phi$ is the angle between the transverse polarization direction
and the normal to the $\ell^+\ell^-$ plane; when integrating 
over all final angular distributions of the $\ell^+\ell^-$ pair -- as we do 
by looking only at variables (\ref{var}) -- the contribution of 
Eq. (\ref{andan}) vanishes. 
 
There exist in the literature other mechanisms to generate SSA in 
Drell-Yan processes \cite{hts,bmt,bl,bq}, based on higher twist quark-gluon 
correlation functions in a generalized pQCD factorization theorem \cite{qs}.
All of them lead to expressions of $A_N$ depending on the angles between
the polarization direction and the final lepton pair plane \cite{cs}, which 
require observation of these angles to be detected and vanish upon 
integration.

In the model of Ref. \cite{meng} a SSA for the differential cross-section
(\ref{udy2}) is computed: in that model a non vanishing asymmetry is 
achieved by introducing orbital angular momentum and surface effects in 
the distribution of valence quarks, resulting in somewhat {\it ad hoc} 
correlations between their polarization and $\bfk_\perp$ distribution. 
It might be that the mechanism is somehow related to the Sivers effect,
although the issue should be further investigated.   

In Refs. \cite{bro1} and \cite{bro2} -- which are meant to be and provide 
important pedagogical examples of SSA -- the processes considered are 
somehow academic and related by crossing. They are respectively 
$\gamma^* \, \pup \to q \, (q \bar q)_0$ 
and $\bar q \, \pup \to \gamma^* \, (q \bar q)_0$ where $(q \bar q)_0$ is 
a spectator scalar diquark. The SSA obtained in these two cases turn out to be
opposite, as predicted by Collins \cite{col2}.  

\vspace{12pt}
\goodbreak
\nd
{\bf 2.2 A simple analytical model}
\nobreak
\vspace{4pt}
\nobreak

In order to give numerical estimates in the next Section, we introduce here 
a simple model for the Sivers asymmetry (\ref{delf1}), and for the
unpolarized distributions, which is similar to the one introduced for 
the polarizing fragmentation function in Ref. \cite{noi2} and has the 
advantage of giving analytically integrable expressions for $A_N$. Such a
model shows explicitely how the asymmetry originates and depends on
$M$, $y$ and $\bfq_T$.  

Let us start from the most general expression for the number density  
of quarks $q$, inside a proton with transverse polarization $\bfP$ and 
three-momentum $\bfp$; the quark has a transverse momentum $\bfk_{\perp}$
and its polarization is not observed. One has 
\be
\hat f_{q/\pup}(x, \bfk_{\perp}) = \hat f_{q/p}(x, k_{\perp})
+ \frac{1}{2} \, \Delta^N f_{q/\pup}(x, k_{\perp}) \> \hat{\bfP} \cdot
\hat{\bfp} \times \hat{\bfk}_{\perp}\>, \label{polden}
\ee
consistently with Eq. (\ref{delf1})
\bea
\!\!\!\!\!\!\!\!\!\!
\hat f_{q/\pup}(x, \bfk_{\perp}) + \hat f_{q/\pdown}(x, \bfk_{\perp})
\!\!\! &=& \!\!\! 2\,\hat f_{q/p}(x, \bfk_{\perp})
=  2\,\hat f_{q/p}(x, k_{\perp}) \label{f+f} \\
\!\!\!\!\!\!\!\!\!\!
\hat f_{q/\pup}(x, \bfk_{\perp}) - \hat f_{q/\pdown}(x, \bfk_{\perp}) 
\!\!\! &=& \!\!\! \Delta^N f_{q/\pup}(x, \bfk_{\perp}) =
\Delta^N f_{q/\pup}(x, k_{\perp}) \> \hat{\bfP} \cdot
\hat{\bfp} \times \hat{\bfk}_{\perp}\,. \label{f-f}
\eea
With the configuration of Fig. 1 one simply has
$\hat{\bfP} \cdot \hat{\bfp} \times \hat{\bfk}_\perp =
(\hat{\bfk}_\perp)_x=\cos\phi_{k_\perp}$.

The Sivers function $\Delta^N f_{q/\pup}(x, k_{\perp})$ must obey the 
positivity bound:
\be
\frac{|\Delta^N f_{q/\pup}(x, k_{\perp})|}
{2 \,\hat f_{q/p}(x, k_{\perp})}  \leq 1
\quad\quad \forall\, x,\,k_{\perp} \,.
\label{posb}
\ee
   
We consider simple factorized and Gaussian forms:
\be
\hat f_{q/p}(x, k_\perp) = f_{q/p}(x)\,g(k_\perp) =
f_{q/p}(x) \,\frac{\beta^2}{\pi}\, e^{-\beta^2 \, k_\perp^2} \>, 
\label{modu}
\ee
\be
\Delta^Nf_{q/\pup}(x, k_\perp) =
\Delta^Nf_{q/\pup}(x)\, h(k_\perp)\>.
\label{modd}
\ee
Notice that $\beta$ can depend on $x$; we have assumed the same 
$k_\perp$ dependence for all flavours, as intrinsic $\bfk_\perp$
of quarks originates via confinement and gluon emissions, which should
be flavour independent processes. Notice also that $\beta^2 = 1/\langle
k_\perp^2 \rangle$ and that $\int d^2\bfk_\perp \, \hat f_{q/p}(x, k_\perp)
= f_{q/p}(x)$. 
 
In order to obviously satisfy the bound (\ref{posb}) we write  
\be
\Delta^Nf_{q/\pup}(x) = 2\,{\mathcal N}_q(x)\,f_{q/p}(x)\,
\label{dnf}
\ee
\be
h(k_\perp)={\mathcal H}(k_\perp)\,g(k_\perp)\,.
\label{hk}
\ee
so that Eq. (\ref{posb}) simply becomes 
\be
|{\mathcal N}_q(x) \, {\mathcal H}(k_\perp)| \le 1
\quad\quad \forall\, x,\,k_\perp \,. \label{bound2}
\ee

We actually impose ${\mathcal N}$ and ${\mathcal H}$ to be separately 
smaller than 1 in magnitude, by choosing simple analytical functional 
forms and dividing each of them by its maximum value: 
\be
{\mathcal N}_q(x) = N_q\,x^{a_q}(1-x)^{b_q}\,
\frac{(a_q+b_q)^{(a_q+b_q)}}{a_q^{a_q}\,b_q^{b_q}}\,,\quad
|N_q|\leq 1\,
\label{nqx}
\ee
\be
{\mathcal H}(k_\perp)=\sqrt{2\,e\,(\alpha^2-\beta^2)}\,
k_\perp\,\exp\,\left[\,-(\alpha^2-\beta^2)\,k_\perp^2\,\right]\,
, \quad \alpha > \beta\>.
\label{hhk}
\ee

Eqs. (\ref{modu})-(\ref{hk}) and (\ref{hhk}) imply:
\be
\Delta^Nf_{q/\pup}(x, k_\perp) = 2 \,  {\mathcal N}_q(x) \,
f_{q/p}(x) \, \frac{\beta^2}{\pi} \, \sqrt{2\,e\,(\alpha^2 - \beta^2)}\,
k_\perp \, e^{-\alpha^2 k_{\perp}^2} \>. \label{del2}
\ee

Inserting the above choice of $\Delta^Nf(x, \bfk_\perp)$ and 
$\hat f(x, \bfk_\perp)$ into Eq. (\ref{ann}) one can perform analytical
integrations. The numerator of Eq. (\ref{ann}) results as:
\bea
N(A_N) &=& \frac{1}{\pi} \, \frac{\beta^3 \bar{\beta}^4}
{(\alpha^2 + \bar{\beta}^2)^2} \,
\left(2\,e\,\frac{\alpha^2-\beta^2}{\beta^2}\right)^{1/2}\,
(\bfq_T)_x \, \exp \left[ - \frac{\alpha^2\bar{\beta}^2}
{\alpha^2 + \bar{\beta}^2} \, q_T^2 \right] \nonumber\\
&\times& \sum_q e_q^2 \, \Delta^Nf_{q/\pup}(x_q) \,
f_{\bar q/p}(x_{\bar q})
\eea
where $\alpha = \alpha(x_q), \, \beta = \beta(x_q)$,
\, $\bar{\beta} = \beta(x_{\bar q})$ with $x_q = 
M \, e^y/\sqrt s, \, x_{\bar q} = M \, e^{-y}/\sqrt s$.
The denominator is
\be
D(A_N) = \frac{1}{\pi} \, \frac{\beta^2 \bar{\beta}^2}{\beta^2 +
\bar{\beta}^2}\exp \left[ - \frac{\beta^2\bar{\beta}^2}
{\beta^2 + \bar{\beta}^2} \, q_T^2 \right] \>
2 \sum_q e_q^2 \, f_{q/p}(x_q) \, f_{\bar q/p}(x_{\bar q}) \>,
\ee
where again $\beta$ stands for $\beta(x_{q})$ and $\bar{\beta}$ for
$\beta(x_{\bar q})$.

The asymmetry (\ref{ann}) in this simple case reads:
\bea
A_N(M,y,\bfq_T) &=&
\beta\,\bar{\beta}^2\>
\frac{\beta^2 + \bar{\beta}^2}{(\alpha^2 + \bar{\beta}^2)^2} \>
\left(2\,e\,\frac{\alpha^2-\beta^2}{\beta^2}\right)^{1/2}\nonumber\\
&\times&\>
(\bfq_T)_x \> \exp \left[ -\left( \frac{\alpha^2}{\alpha^2 +
\bar{\beta}^2} - \frac{\beta^2}{\beta^2 + \bar{\beta}^2} \right)
\bar{\beta}^2 \, q_T^2 \right] \nonumber \\ 
&\times& 
\frac{1}{2}\>
\frac{\sum_q e_q^2 \, \Delta^Nf_{q/\pup}(x_q) \, f_{\bar q/p}(x_{\bar q})} 
{\sum_q e_q^2 \, f_{q/p}(x_q) \, f_{\bar q/p}(x_{\bar q})} \> \cdot
\label{anng}
\eea
For a generic configuration, different from the one of Fig. 1, one
simply replaces $(\bfq_T)_x$ with 
$\hat{\bfP} \cdot \hat{\bfp} \times \bfq_T$.
 
When, as it is often the case in the literature, $\alpha$ and $\beta$ are 
taken to be independent of $x$, Eq. (\ref{anng}) simplifies to:
\bea
A_N(M,y,\bfq_T) &=&
\frac{2\,\beta^5}{(\alpha^2 + \beta^2)^2} \> 
\left(2\,e\,\frac{\alpha^2-\beta^2}{\beta^2}\right)^{1/2}\nonumber\\
&\times&\>(\bfq_T)_x \> \exp \left[ -\left( \frac{\alpha^2}{\alpha^2 + 
\beta^2} - \frac{1}{2} \right) \beta^2 \, q_T^2 \right] 
\nonumber \\ 
&\times&\> 
\frac{1}{2}\>\frac{\sum_q e_q^2 \, \Delta^Nf_{q/\pup}(x_q) \,
f_{\bar q/p}(x_{\bar q})} 
{\sum_q e_q^2 \, f_{q/p}(x_q) \, f_{\bar q/p}(x_{\bar q})} \> \cdot
\label{anng0}
\eea

The same simplified expression holds also at $y=0$, where  
one has $x_q = x_{\bar q} = M/\sqrt s$.
 
\vspace{18pt}
\goodbreak
\nd
{\bf 3. SSA in Drell-Yan processes, numerical estimates}
\nobreak
\vspace{6pt}
\nobreak

In this Section we shall present numerical estimates of the SSA for the 
Drell-Yan process, originating from Sivers effect, in kinematical 
configurations relevant for RHIC experiments.

We follow the simple analytical model discussed in the previous Section; 
we need to fix the parameters of the model, that is the functions $\beta(x)$,
$\alpha(x)$ and the flavour-dependent coefficients $N_q$, $a_q$, $b_q$
appearing respectively in Eqs. (\ref{modu}), (\ref{del2}) and (\ref{nqx}).
The unpolarized and $\bfk_{\perp}$ integrated partonic distributions are 
chosen among the sets available in the literature; in particular, we adopt 
the GRV94 set \cite{grv94}.

Some information on $\beta$, entering the $\bfk_\perp$ dependence in the 
unpolarized parton distributions, can be obtained by considering the
available data on the unpolarized cross-sections for inclusive particle 
production processes. This not only supplies information on $\beta$, but 
allows a crucial consistency check of our formalism, as it shows whether 
or not we can reproduce the unpolarized data.  

Experimental data are presently available for pion, prompt photon production 
and for Drell-Yan processes, in different kinematical configurations.
As mentioned in the introduction, several papers \cite{pkt} have studied
in details the unpolarized cross-sections for these processes within 
perturbative QCD at LO and NLO, emphasizing the crucial role of 
intrinsic transverse momentum.

We have performed an independent attempt of reproducing the available
data within the same approach utilized in the calculation of the SSA.
Since SSA have been studied at present only at leading order and 
leading twist, the same level of accuracy has been adopted for the unpolarized 
cross-sections, keeping in mind that there might be NLO corrections 
(the so-called $K$ factors) the value of which may vary, depending on the 
process and on the kinematical configuration considered,
approximately between a factor 1 and 3.

Our aim is not to reproduce as accurately as possible the unpolarized 
cross-sections, as in \cite{pkt}, but rather to show that within the same 
approach used for the SSA it is possible to reproduce the unpolarized
cross-section values, up to an overall factor between 1 and 3, due to NLO 
corrections, scale-dependences, etc. These corrections should have little 
effect on the SSA, largely canceling in the ratio of cross-sections.

A full account of the combined study of unpolarized cross-sections and SSA
in different processes is outside the scope of this paper, and will be given
in a separate forthcoming paper \cite{DM}.
It suffices to state here that, as a result of this comprehensive analysis, a 
value of $\beta$ independent of $x$, $\beta=1.25\,({\rm GeV}/c)^{-1}$, 
corresponding to $\sqrt{\langle\,k_\perp^2\,\rangle} = 0.8\,{\rm GeV}/c$, 
allows a good description of the unpolarized processes and is appropriate 
for our scope. This value is in good agreement with the results of the 
abovementioned papers devoted to unpolarized cross-section calculations 
\cite{pkt}, and will be adopted in the sequel.

One might argue that the value of $\beta$ varies with energy and that RHIC 
works at c.m. energies much larger than most of previous experiments;
this could be tested when RHIC data on unpolarized cross-sections will become 
available and, for the moment, we keep using the value
$\beta=1.25\, ({\rm GeV}/c)^{-1}$, which corresponds to an already large
$\langle\,k_\perp\,\rangle$ value. We only notice that in our approach 
changes in $\beta$ should influence only the dependence of the SSA on 
$|\bfq_T|$. In the following, we will show such a dependence by presenting some
of our results for $\beta = 0.83 \,({\rm GeV}/c)^{-1}$, which means
$\sqrt{\langle\,k_\perp^2\,\rangle} = 1.2\,{\rm GeV}/c$. 

Adopting a $\beta$ parameter independent of $x$ and, as we do, of quark 
flavour, simplifies our analytical expressions as one gets 
$\bar{\beta}\equiv\beta$, which allows the use of Eq.~(\ref{anng0}).
% for all values of the rapidity $y$, rather than only for $x_{\bar q} = x_q, \> y=0$.

As a second step, we must fix the parameter $\alpha$ determining the gaussian
shape of the Sivers function, see Eq. (\ref{del2}). Since the positivity 
bound, Eq.~(\ref{posb}), requires $\alpha > \beta$, one can write 
$\alpha^2=\beta^2/r$, where $r$ is a numerical parameter in the range $(0,1)$. 
A similar procedure was already followed by us in a previous work on
transverse hyperon polarization \cite{noi2}, and an optimal guess
for the parameter $r$ was found, $r\simeq 0.7$, value which we also
adopt here. The same value allows a good description, using Sivers function, 
of the E704 data on $A_N$ \cite{e704,DM}.

The next step is to obtain information on the
$x$-dependent part of the Sivers function, $\Delta^Nf_{q/\pup}(x)$.
Indications on the values of the parameters entering Eq.~(\ref{nqx}) can 
be obtained by exploiting the data on SSA measured by the E704 experiment 
\cite{e704} for the process $\pup\,p\to\pi\,X$.

Estimates of the Sivers function from E704 results were already given
and used to make predictions for SSA in similar processes in \cite{nois}.
In those first papers, however, a number of simplifying approximations were 
adopted which are not kept in the present, more general, analysis. Those 
assumptions amounted to consider $\bfk_\perp$ effects only minimally,
in places where their neglect would lead to vanishing results; also,
no gaussian $k_\perp$ distribution was introduced, but rather a simple
$\delta$-like dependence which leads to the effective use of an average  
$k_\perp$ in a simplified (planar) kinematics.  

Although the simplified procedure might be a good approximation for 
single spin asymmetries, which are ratios of cross-sections, 
a more general and refined analysis, using full $\bfk_\perp$ dependences 
according to Eq. (\ref{ltgen}), is in progress and a detailed 
analysis will be presented elsewhere \cite{DM}; we only state here that 
good reproductions of both the cross-sections and the E704 asymmetry 
are possible and a good set of values for the parameters $N_q$, $a_q$
and $b_q$ of the Sivers function is: 
\bea
N_u &=&  \phantom{-}0.5 \qquad a_u = 2.0 \qquad b_u = 0.3 \qquad,\nonumber\\
N_d &=& -1.0 \qquad a_d = 1.5 \qquad b_d = 0.2 \qquad .
\label{nab}
\eea
Notice that we are assuming that only valence quarks in the proton
give a nonvanishing Sivers function. 

The above values allow a good description of the E704 data on the SSA in 
the $\pup \, p \to \pi \, X$ process \cite{e704}, assuming that $A_N$ is 
totally generated by Sivers asymmetric distribution. We use them here in order
to estimate $A_N$ for a D-Y process at a much higher energy: we are well
aware that this might be a rough extrapolation, as we neglect pQCD evolution,
other possible mechanisms, and so on. However, since information on Sivers 
function is so scarce, even rough estimates of the SSA in Drell-Yan processes 
are important and useful, and we shall proceed with our program, keeping in 
mind all these remarks. 

One further uncertainty concerns the sign of the asymmetry: as noticed 
by Collins \cite{col2} and checked in Ref. \cite{bro2}, the Sivers asymmetry 
has opposite signs in Drell-Yan and SIDIS, respectively related 
to $s$-channel and $t$-channel elementary reactions. As in $p-p$ 
interactions we expect, at large $x_F$, a dominant contribution
from $t$-channel quark processes, we think that the Sivers function
extracted from $p-p$ data should be opposite to that contributing
to D-Y processes. Our numerical estimates will then be given with the same
parameters as in Eq.~(\ref{nab}), {\it changing the signs of} $N_u$ and $N_d$. 
Given these considerations, even a simple comparison of the sign of our
estimates with data might be significant.

We have thus completed the choice of parameters of our model, and we are now 
able to give predictions for the Drell-Yan process, without forgetting all
the cautious comments made above.

We only add that a different modelization of the Sivers function, originally 
proposed by Collins for his function \cite{col} and often adopted in the 
literature, could be used. This functional form does not allow to perform 
exact analytical integrations on transverse momentum, and we prefer using the 
form proposed here. However, we have checked that for consistent choices of 
the parameters, the two parameterizations give very similar results, since 
they mainly differ at relatively large transverse momentum, where the gaussian 
dumping is already effective.

An important point has to be noticed regarding the parameterization of the
Sivers function. In order to reproduce the E704 results, it has to have
a valence-like behaviour. That is,
$\Delta^Nf_{q/p^{\uparrow}}(x)/f_{q/p}(x) \sim x^a$, where $a>0$,
for small $x$.

This is a very general feature, somehow required by the experimental data,
which may have strong effects for the SSA in Drell-Yan processes,
leading to tiny values of $A_N$ when the $x$ values explored are
relatively small. 
This prediction and its extent of validity can be tested at RHIC.
In what follows, we shall show some estimates for the SSA in realistic 
kinematical regions:
\be
6\, {\rm GeV} \leq M \leq 10\, {\rm GeV};\qquad\qquad -2 \leq y \leq 2 \>.
\ee

To show the dependence on the range of invariant mass covered,
we will consider also the alternative set of kinematical cuts

\be
10\, {\rm GeV} \leq M \leq 20\, {\rm GeV};\qquad\qquad -2 \leq y \leq 2 \>.
\ee

These kinematical regions should be easily explored in RHIC experiments,
while assuring that the Drell-Yan process is the dominating
contribution to lepton pair production. Notice that with the above cuts the 
lowest possible value of $x$ reached is around $10^{-3}$.

In our simple analytical model,
%due to the neglecting of terms of the order $k_{\perp}/M$
%and to the use of an $x$ and flavour-independent $\beta$ factor, 
the dependences of $A_N$ on $\bfq_T$ and $x_q$, $x_{\bar q}$ (or $M$, $y$) 
are completely uncorrelated, and may be treated separately.
In fact, assuming $\bar{\beta}\equiv \beta$ and $\alpha^2=\beta^2/r$,
Eq.~(\ref{anng}) now reads

\bea
A_N(M,y,\bfq_T) &=&
{\mathcal Q}(q_T,\phi_{q_{_T}})\>{\mathcal A}(M,y) \nonumber\\
&=&2\,\frac{r^2}{(1+r)^2}\>\left(\,2\,e\,\frac{1-r}{r}\,\right)^{1/2}\>
\beta \,q_T \,\cos\phi_{q_{_T}}\>\exp\,\left[\,-\frac{1}{2}\,\frac{1-r}{1+r}\,
\beta^2\,q_T^2\,\right]\nonumber\\
&\times&\>\frac{1}{2}\>\frac{\sum_q e_q^2 \, \Delta^Nf_{q/\pup}(x_q) \,
f_{\bar q/p}(x_{\bar q})} {\sum_q e_q^2 \, f_{q/p}(x_q) \, 
f_{\bar q/p}(x_{\bar q})} \> \cdot
\label{anr}
\eea
where $\phi_{q_{_T}}$ is the azimuthal angle of $\bfq_T$, as defined 
in Fig. 1.

The second line of the above equation, the function ${\mathcal Q}(\bfq_T)$,
describes the $\bfq_T$ dependence of the asymmetry, which turns out to be 
very simple and directly related to the $\bfk_{\perp}$ behaviour of the 
Sivers function (via the parameter $r$). One can easily see that 
${\mathcal Q}(q_T)$ has a maximum when 
\be
q_T = q_T^M = \sqrt{(1+r)/(1-r)}/\beta \>, \label{max}
\ee
where its value is ${\mathcal Q}(q_T^M)\equiv {\mathcal Q}_M = 
[\,2\,r/(1+r)\,]^{3/2}$. Notice that the position of the maximum
depends on the parameter $\beta$, while ${\mathcal Q}_M$ only
depends on $r=\beta^2/\alpha^2$. In particular, when $r=0.7$,
$q_T^M\simeq 2.38/\beta$, and ${\mathcal Q}_M\simeq 0.75$. 

In order to collect statistical significance, one could integrate the
asymmetry over $\bfq_T$, but this would lead to a vanishing result, due
to the $\cos\phi_{q_{_T}}$ factor. However, one can think of integrating over 
$\phi_{q_{_T}}$ in the range $[0,\pi/2]$ only, or alternatively of taking into 
account the change of sign in the different quadrants.
In both ways, the $\phi_{q_{_T}}$ integration gives
an overall factor $\int\,d\phi\,|\cos\phi|/\int\,d\phi=2/\pi$.

We can then consider the magnitude of the asymmetry averaged
over $\bfq_T$ up to an upper value of $q_T = q_{T1}$. In our simple model,
Eq. (\ref{anr}), one obtains 

\bea
&&\left\langle\, |\,A_N(M,y,\bfq_T)\,|\,\right\rangle_{q_{T1}} = 
\frac{\int^{q_{T1}} d^2 \bfq_T\>|\,A_N(M,y,\bfq_T)\,|
\> d\sigma}{\int^{q_{T1}} d^2 \bfq_T \> d\sigma}
= \frac{\int^{q_{T1}} d^2 \bfq_T \> (d\sigma^\uparrow - d\sigma^\downarrow)}
{\int^{q_{T1}} d^2 \bfq_T \> (d\sigma^\uparrow + d\sigma^\downarrow)}
\nonumber\\
\!\!\!&=&\!\!\! 
\frac{1}{\pi}\>r^{3/2} \left(\,2\,e\,\frac{1-r}{1+r}\,\right)^{1/2}
\left\{\,\sqrt{\pi}\,{\rm Erf}\left( w \right)
-2\,w\,e^{-w^2}\>\right\} \> 
\left(\,1-e^{-(1+r)\,w^2/2}\,\right)^{-1}
% \nonumber \\ \!\!\!&\times&\!\!\!
\>  {\mathcal A}(M,y) \nonumber\\
\!\!\!&=&\!\!\! \tilde{\mathcal Q}(q_{T1})\>{\mathcal A}(M,y)
\label{aveq}
\eea
where $d\sigma$ stands for $d\sigma/dy \, dM^2 \, d^2\bfq_T$,
$w = \beta\,q_{T1}/\sqrt{1+r}$ and ${\mathcal A}(M,y)$ is given by the 
last line of Eq. (\ref{anr}). In particular, when $q_{T1} \to \infty$, 
one finds the very simple result:

\be
\left\langle\, |\,A_N(M,y,\bfq_T)\,|\,\right\rangle_{\infty}
= \left(\,\frac{2\,e}{\pi}\,\right)^{1/2}\>r^{3/2}\>
\left(\,\frac{1-r}{1+r}\,\right)^{1/2}\>{\mathcal A}(M,y) \>.
\label{aveqinf}
\ee
This expression holds with a good accuracy already at $q_{T1} \gsim 1/\beta$.

Notice again that, concerning the intrinsic motion dependence, the above 
average is independent of $\beta$ and depends only on $r$, via a function
which has a maximum at $r=(\sqrt{10}-1)/3\simeq 0.72$, very
close to the value $r=0.7$ adopted in our analysis. 

In Figs. 2 to 5 we show our model estimates for the SSA, starting from 
Eq. (\ref{anr}). In Fig. 2 we plot $A_N$ as a function of $y$: the asymmetry 
is averaged over $M$, in the two kinematical ranges $6\leq M \leq 10$ GeV and 
$10 \leq M \leq 20$ GeV, corresponding to the solid and the dashed curve 
respectively. We have fixed $q_T=q_T^M$, Eq. (\ref{max}), and 
$\phi_{q_{_T}}=0$, which maximizes the $\bfq_T$-dependent part of the 
asymmetry. 

In Fig. 3 we show the same asymmetry, averaged over different rapidity ranges, 
$|y| \leq 2$ (solid curve) and $0 \leq y \leq 2$ (dashed curve). 

We also show the dependence of $A_N$ on $x_F$ in Fig. 4 and on $x_q$ in 
Fig. 5; these partonic variables ($x_F = x_q - x_{\bar q}$) are obviously 
related to $y$ and $M$ via Eq. (\ref{odel}). The integration over the non 
observed variables is done in such a way as to correspond to the   
kinematical ranges $-2 \leq y \leq 2$, $6 \leq M \leq 10$ GeV (solid curve)
and $-2 \leq y \leq 2$, $10 \leq M \leq 20$ GeV (dashed curve). 

All asymmetries in Figs. 2-5 are evaluated at $q_T=q_T^M$: the value of the 
SSA at different $q_T$ values may be obtained by simply rescaling the results 
of Fig.s 2-5 by the factor ${\mathcal Q}(q_T)/{\mathcal Q}(q_T^M)$. 
This factor is plotted in Fig. 6. In order to show how its behaviour 
depends on the parameter $\beta$, we present results for
$\beta=1.25 \, ({\rm GeV}/c)^{-1}$ and $\beta=0.83 \, ({\rm GeV}/c)^{-1}$.

Analogously, one can obtain the magnitude of the asymmetry averaged over
$\bfq_T$, Eq. (\ref{aveq}), by rescaling the values given in Figs. 2-5 by the 
appropriate factor: this factor is obtained by dividing
$\tilde{\mathcal Q}(q_{T1})$, see Eq. (\ref{aveq}),
by ${\mathcal Q}(q_T^M)$, and it is shown in Fig. 7.

Our numerical estimates show that $A_N$ can be well measurable within 
RHIC expected statistical accuracy. The actual values depend on the
assumed functional form of the Sivers function and its role with valence 
quarks only. This reflects in the increase of $A_N$ to sizeable values at 
large $x_q$, Fig. 5, and at large $x_F$, Fig. 4; in fact, $x_F$ large and 
positive implies a large $x_q > x_F$.  

\vspace{18pt}
\goodbreak
\nd
{\bf 4. Comments and conclusions}
\nobreak
\vspace{6pt}
\nobreak

The single transverse spin phenomenology, within QCD dynamics and the
factorization scheme, is a rich and interesting subject. It combines 
simple pQCD spin dynamics with new long distance properties of quark
distribution and fragmentation; the experimental measurements are relatively
easy and clear, many have been and many more will be performed in the 
near future, both at nucleon-nucleon and lepton-nucleon facilities. 

Very recently a large single transverse spin asymmetry has been observed
at RHIC, at the very first spin measurement at $\sqrt s =$ 200 GeV, in 
$\pup \, p \to \pi \, X$ processes \cite{bnl}; despite the large energy the 
asymmetry is not negligible, contrary to naive expectations. 

The approach adopted here - and in many previous papers - requires the 
explicit control of parton intrinsic motion, which cannot be simply 
integrated over in the different non perturbative functions and the elementary
dynamics, but must be kept into account wherever it can give new effects.
Actually, this is true also for a correct computation of the unpolarized 
cross-sections, as it has been known for a long time, although somewhat 
forgotten. When dealing with spin dependences, the intrinsic motion is a
rich and unexpected source of many new effects.        

We have presented here the explicit formalism for computing single transverse 
spin asymmetries in Drell-Yan processes, within a generalized QCD 
factorization theorem formulated with $\bfk_\perp$ dependent distribution
functions. Simple gaussian forms have been assumed and available data from 
other processes have been exploited, in order to give estimates for single
spin effects in D-Y production at RHIC, which should be of interest for the 
incoming measurements. Again, sizeable and measurable values have been found.  

Our approach can and will be extended to the study of any $A \, B \to C \, X$ 
process, at large energy and moderate to large momentum transfer; the parton
intrinsic motion is relevant both in polarized and unpolarized processes and
the resulting phenomenology might explain or anticipate many subtle and 
unexpected results.    

\vskip 18pt
\goodbreak
\nd
{\bf Acknowledgements}
\vskip 6pt
We would like to thank D. Boer for many useful discussions;
U.D. and F.M. thank COFINANZIAMENTO MURST-PRIN for partial support.

\normalsize

\newpage

\noindent {\bf Figure captions}

\vspace{12pt}

\noindent{\bf Fig. 1:\ }
The kinematical configuration considered in this paper. The $\gamma^*$
four-momentum defines all our observables; the dependence on the angle
between the $\gamma^*$-$z$ and the $\gamma^*$-$(\ell^+\ell^-)$ planes is
integrated over in Eq.s~(\ref{ddy2}) and (\ref{udy2}).

\vspace{8pt}

\noindent{\bf Fig. 2:\ }
The single spin asymmetry $A_N$ for the Drell-Yan process, see
Eq.~(\ref{anr}), at RHIC energies, $\sqrt{s}=200$ GeV, as a function of
the rapidity $y$ and averaged over the invariant mass ranges
$6\leq M\leq 10$ GeV (solid line) and $10 \leq M \leq 20$ GeV (dashed line).
Results are given at $q_T=q_T^M$, see Eq.~(\ref{max}),
and $\phi_{q_{_T}}=0$, which maximizes the effect.
Furthermore, we have used $r=0.7$, the parameters of Eq.~(\ref{nab})
for the Sivers function (see text for further details) and the
parameterization GRV94 \cite{grv94} for the unpolarized parton distributions.
Notice that the asymmetry is practically negligible in the range $y<0$.

\vspace{8pt}

\noindent{\bf Fig. 3:\ }
The single spin asymmetry $A_N$ for the Drell-Yan process, see
Eq.~(\ref{anr}),  at RHIC energies, $\sqrt{s}=200$ GeV, as a function of
the invariant mass $M$ and averaged over the rapidity ranges
$-2 \leq y\leq 2$ (solid line) and $0\leq y\leq 2$ (dashed line).
Results are given at $q_T=q_T^M$, see Eq.~(\ref{max}),
and $\phi_{q_{_T}}=0$, which maximizes the effect.
Furthermore, we have used $r=0.7$, the parameters of Eq.~(\ref{nab})
for the Sivers function (see text for further details) and the
parameterization GRV94 \cite{grv94} for the unpolarized parton distributions.

\vspace{8pt}

\noindent{\bf Fig. 4:\ }
The single spin asymmetry $A_N$ for the Drell-Yan process, see
Eq.~(\ref{anr}),  at RHIC energies, $\sqrt{s}=200$ GeV, as a function of
the Feynman variable $x_F=x_q-x_{\bar{q}}$ and averaged over the
rapidity and the invariant mass in the ranges $-2\leq y\leq 2$,
$6\leq M\leq 10$ GeV (solid line) and $-2\leq y\leq 2$,
$10\leq M\leq 20$ GeV (dashed line).
Results are given at $q_T=q_T^M$, see Eq.~(\ref{max}),
and $\phi_{q_{_T}}=0$, which maximizes the effect.
Furthermore, we have used $r=0.7$, the parameters of Eq.~(\ref{nab})
for the Sivers function (see text for further details) and the
parameterization GRV94 \cite{grv94} for the unpolarized parton distributions.
The two curves almost coincide but the solid line, corresponding to
a lower invariant mass range, cannot reach values of $x_F\gsim 0.36$
within the given $y$ range.   
Notice that the asymmetry is practically negligible in the range $x_F<0$.

\vspace{8pt}

\noindent{\bf Fig. 5:\ }
The single spin asymmetry $A_N$ for the Drell-Yan process, see
Eq.~(\ref{anr}),  at RHIC energies, $\sqrt{s}=200$ GeV, as a function of
$x_q$ and averaged over the
rapidity and the invariant mass in the ranges $-2\leq y\leq 2$,
$6\leq M\leq 10$ GeV (solid line) and $-2\leq y\leq 2$,
$10\leq M\leq 20$ GeV (dashed line).
Results are given at $q_T=q_T^M$, see Eq.~(\ref{max}),
and $\phi_{q_{_T}}=0$, which maximizes the effect.
Furthermore, we have used $r=0.7$, the parameters of Eq.~(\ref{nab})
for the Sivers function (see text for further details) and the
parameterization GRV94 \cite{grv94} for the unpolarized parton distributions.
The two curves almost coincide but the solid line, corresponding to
a lower invariant mass range, cannot reach values of $x_q\gsim 0.37$
within the given $y$ range.   

\vspace{8pt}

\noindent{\bf Fig. 6:\ }
The factor ${\mathcal Q}(q_T)/{\mathcal Q}(q_T^M)$, see
Eq.s~(\ref{anr}) and (\ref{max}), plotted as a function of $q_T$,
for $\beta=1.25$ (GeV/$c)^{-1}$ (solid line) and
$\beta=0.83$ (GeV/$c)^{-1}$ (dashed line), corresponding respectively
to $\langle\,k_\perp^2\,\rangle^{1/2}=0.8$ GeV$/c$ and
$\langle\,k_\perp^2\,\rangle^{1/2}=1.2$ GeV$/c$.
This factor can be used to rescale the asymmetries given
in Fig.s 2-5, at $q_T=q_T^M$, to their values at $q_T$ 
different from $q_T^M$.

\vspace{8pt}

\noindent{\bf Fig. 7:\ }
The factor $\tilde{\mathcal Q}(q_{T1})/{\mathcal Q}(q_T^M)$,
see Eq.s~(\ref{aveq}) and (\ref{max}), plotted as a function of $q_{T1}$,
for  $\beta=1.25$ (GeV/$c)^{-1}$ (solid line) and
$\beta=0.83$ (GeV/$c)^{-1}$ (dashed line), corresponding respectively
to $\langle\,k_\perp^2\,\rangle^{1/2}=0.8$ GeV$/c$ and
$\langle\,k_\perp^2\,\rangle^{1/2}=1.2$ GeV$/c$.
This factor can be used to obtain from the asymmetries given in Fig.s 2-5
(at fixed $\bfq_T$, $q_T=q_T^M$ and $\phi_{q_{_T}}=0$)
the corresponding asymmetries averaged over $\bfq_T$ up to
$|\bfq_T|=q_{T1}$ (see text for further details).

\newpage

%FIG1
\begin{figure}[ht]
\begin{center}
\hspace*{0.7cm}
\mbox{~\epsfig{file=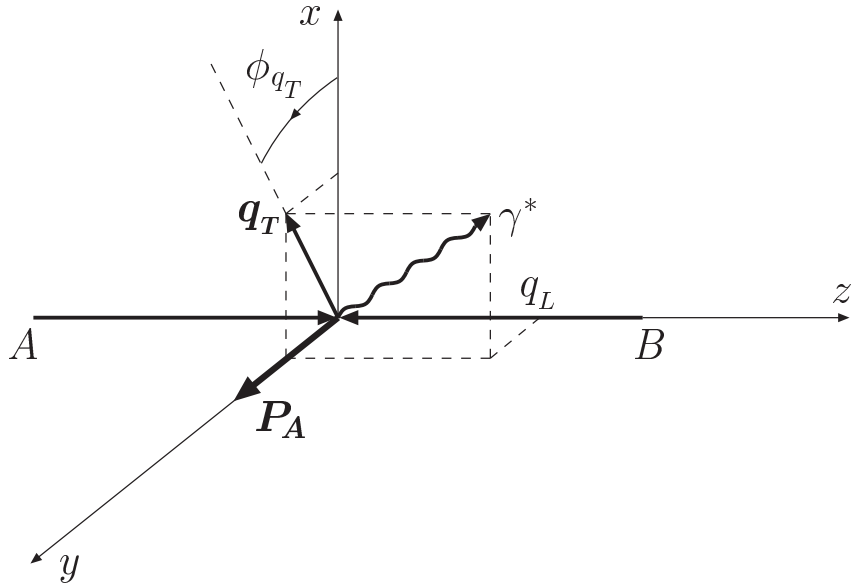,angle=0,width=.9\textwidth}}
\caption[a1]{\small{
The kinematical configuration considered in this paper. The $\gamma^*$
four-momentum defines all our observables; the dependence on the angle
between the $\gamma^*$-$z$ and the $\gamma^*$-$(\ell^+\ell^-)$ planes is
integrated over in Eq.s~(\ref{ddy2}) and (\ref{udy2}).
}}
\end{center}
\end{figure}

\newpage

%FIG2
\begin{figure}[ht]
\begin{center}
\hspace*{1cm}
\mbox{~\epsfig{file=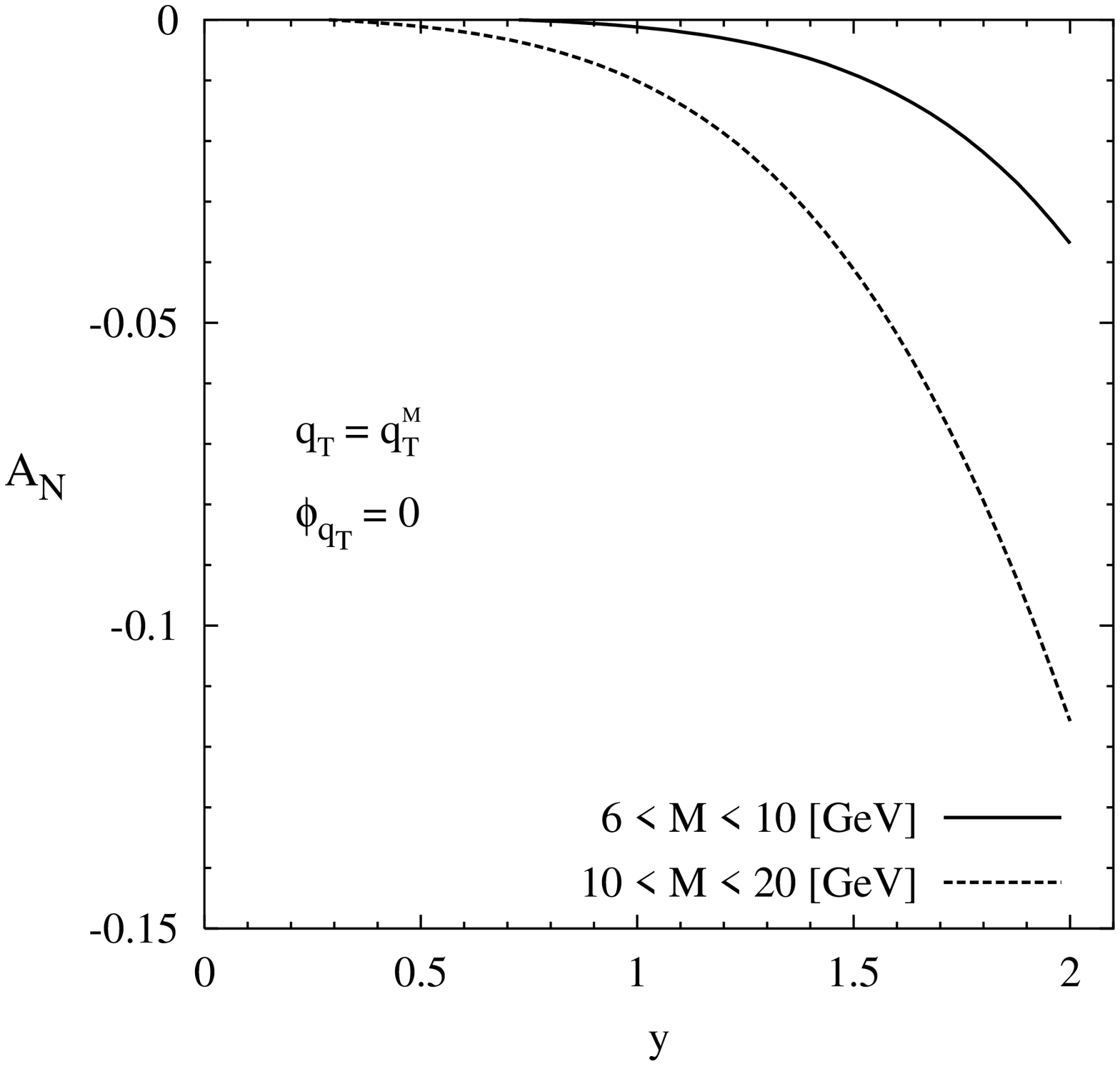,angle=-90,width=.9\textwidth}}
\caption[a2]{\small{
The single spin asymmetry $A_N$ for the Drell-Yan process, see
Eq.~(\ref{anr}), at RHIC energies, $\sqrt{s}=200$ GeV, as a function of
the rapidity $y$ and averaged over the invariant mass ranges
$6\leq M\leq 10$ GeV (solid line) and $10 \leq M \leq 20$ GeV (dashed line).
Results are given at $q_T=q_T^M$, see Eq.~(\ref{max}),
and $\phi_{q_{_T}}=0$, which maximizes the effect.
Furthermore, we have used $r=0.7$, the parameters of Eq.~(\ref{nab})
for the Sivers function (see text for further details) and the
parameterization GRV94 \cite{grv94} for the unpolarized parton distributions.
Notice that the asymmetry is practically negligible in the range $y<0$.
}}
\end{center}
\end{figure}

\newpage

%FIG3
\begin{figure}[ht]
\begin{center}
\hspace*{1cm}
\mbox{~\epsfig{file=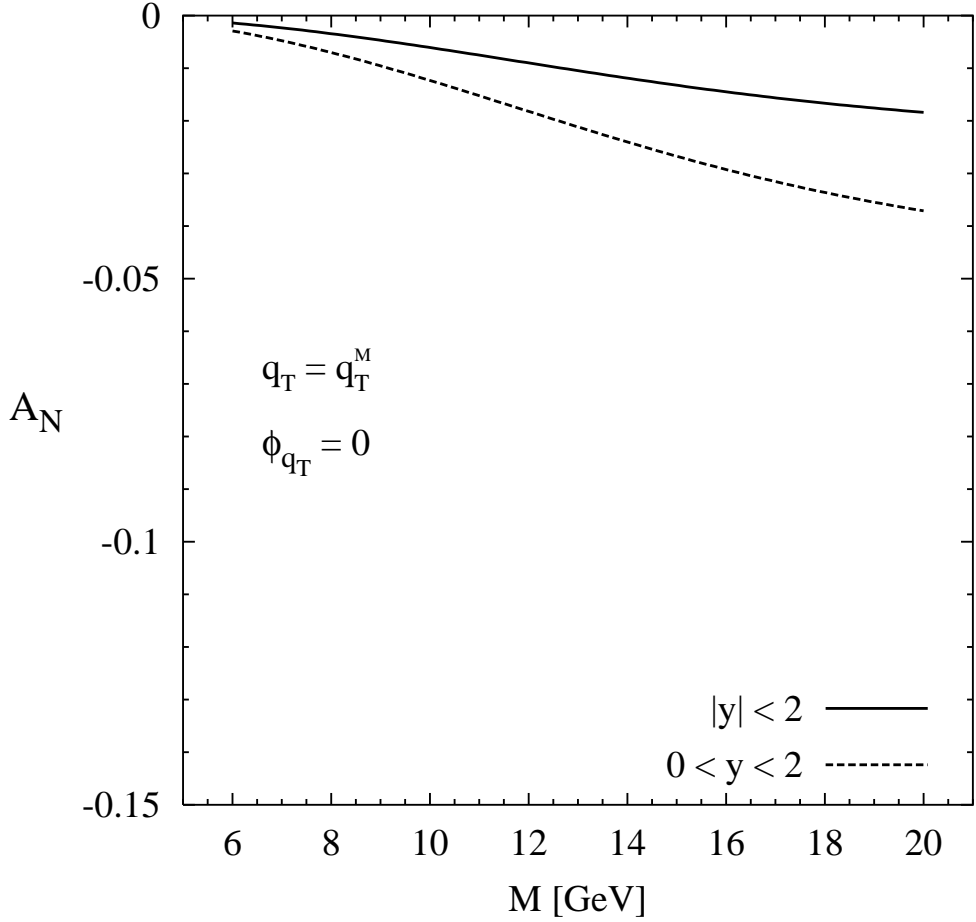,angle=-90,width=.9\textwidth}}
\caption[a3]{\small{
The single spin asymmetry $A_N$ for the Drell-Yan process, see
Eq.~(\ref{anr}),  at RHIC energies, $\sqrt{s}=200$ GeV, as a function of
the invariant mass $M$ and averaged over the rapidity ranges
$-2 \leq y\leq 2$ (solid line) and $0\leq y\leq 2$ (dashed line).
Results are given at $q_T=q_T^M$, see Eq.~(\ref{max}),
and $\phi_{q_{_T}}=0$, which maximizes the effect.
Furthermore, we have used $r=0.7$, the parameters of Eq.~(\ref{nab})
for the Sivers function (see text for further details) and the
parameterization GRV94 \cite{grv94} for the unpolarized parton distributions.
}}
\end{center}
\end{figure}

\newpage

%FIG4
\begin{figure}[ht]
\begin{center}
\hspace*{1cm}
\mbox{~\epsfig{file=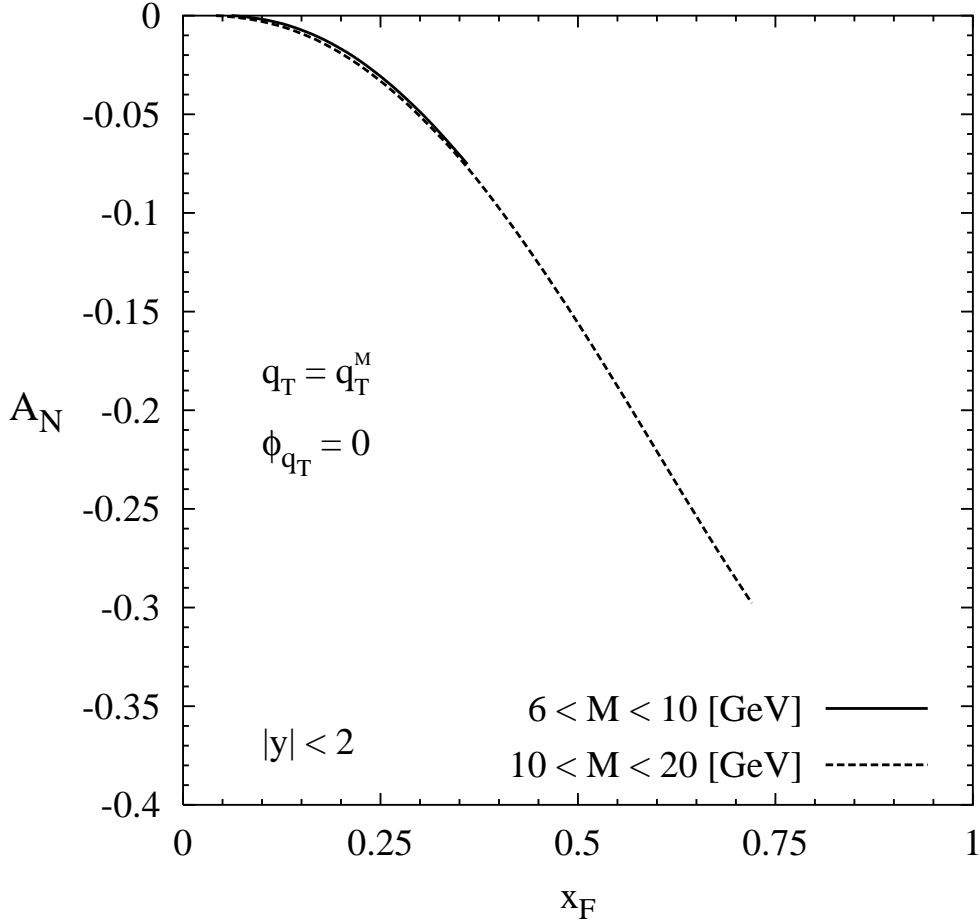,angle=-90,width=.9\textwidth}}
\caption[a4]{\small{
The single spin asymmetry $A_N$ for the Drell-Yan process, see
Eq.~(\ref{anr}),  at RHIC energies, $\sqrt{s}=200$ GeV, as a function of
the Feynman variable $x_F=x_q-x_{\bar{q}}$ and averaged over the
rapidity and the invariant mass in the ranges $-2\leq y\leq 2$,
$6\leq M\leq 10$ GeV (solid line) and $-2\leq y\leq 2$,
$10\leq M\leq 20$ GeV (dashed line).
Results are given at $q_T=q_T^M$, see Eq.~(\ref{max}),
and $\phi_{q_{_T}}=0$, which maximizes the effect.
Furthermore, we have used $r=0.7$, the parameters of Eq.~(\ref{nab})
for the Sivers function (see text for further details) and the
parameterization GRV94 \cite{grv94} for the unpolarized parton distributions.
The two curves almost coincide but the solid line, corresponding to
a lower invariant mass range, cannot reach values of $x_F \gsim 0.36$
within the given $y$ range.
Notice that the asymmetry is practically negligible in the range $x_F<0$.
}}
\end{center}
\end{figure}

\newpage

%FIG5
\begin{figure}[ht]
\begin{center}
\hspace*{1cm}
\mbox{~\epsfig{file=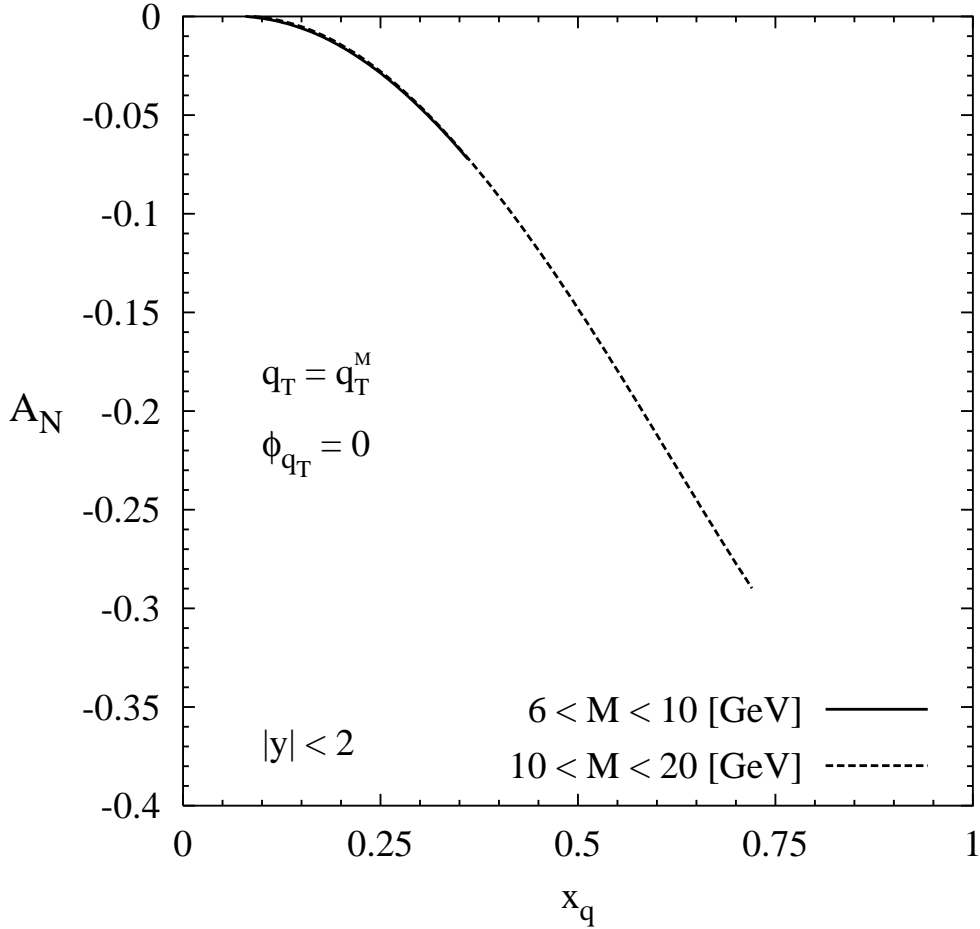,angle=-90,width=.9\textwidth}}
\caption[a5]{\small{
The single spin asymmetry $A_N$ for the Drell-Yan process, see
Eq.~(\ref{anr}),  at RHIC energies, $\sqrt{s}=200$ GeV, as a function of
$x_q$ and averaged over the
rapidity and the invariant mass in the ranges $-2\leq y\leq 2$,
$6\leq M\leq 10$ GeV (solid line) and $-2\leq y\leq 2$,
$10\leq M\leq 20$ GeV (dashed line).
Results are given at $q_T=q_T^M$, see Eq.~(\ref{max}),
and $\phi_{q_{_T}}=0$, which maximizes the effect.
Furthermore, we have used $r=0.7$, the parameters of Eq.~(\ref{nab})
for the Sivers function (see text for further details) and the
parameterization GRV94 \cite{grv94} for the unpolarized parton distributions.
The two curves almost coincide but the solid line, corresponding to
a lower invariant mass range, cannot reach values of $x_q\gsim 0.37$
within the given $y$ range.
}}
\end{center}
\end{figure}

\newpage

%FIG6
\begin{figure}[ht]
\begin{center}
\hspace*{1cm}
\mbox{~\epsfig{file=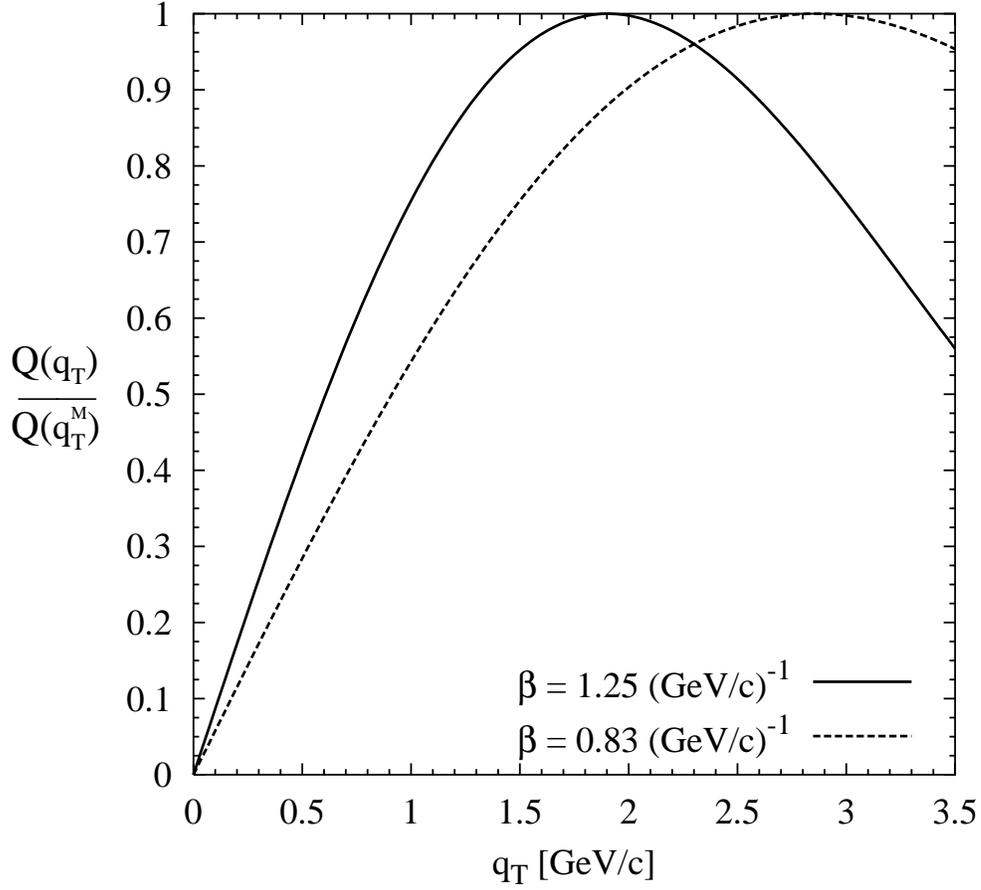,angle=-90,width=.9\textwidth}}
\caption[a6]{\small{
The factor ${\mathcal Q}(q_T)/{\mathcal Q}(q_T^M)$, see
Eq.s~(\ref{anr}) and (\ref{max}), plotted as a function of $q_T$,
for $\beta=1.25$ (GeV/$c)^{-1}$ (solid line) and
$\beta=0.83$ (GeV/$c)^{-1}$ (dashed line), corresponding respectively
to $\langle\,k_\perp^2\,\rangle^{1/2}=0.8$ GeV$/c$ and
$\langle\,k_\perp^2\,\rangle^{1/2}=1.2$ GeV$/c$.
This factor can be used to rescale the asymmetries given
in Fig.s 2-5, at $q_T=q_T^M$, to their values at $q_T$
different from $q_T^M$.
}}
\end{center}
\end{figure}

\newpage

%FIG7
\begin{figure}[ht]
\begin{center}
\hspace*{1cm}
\mbox{~\epsfig{file=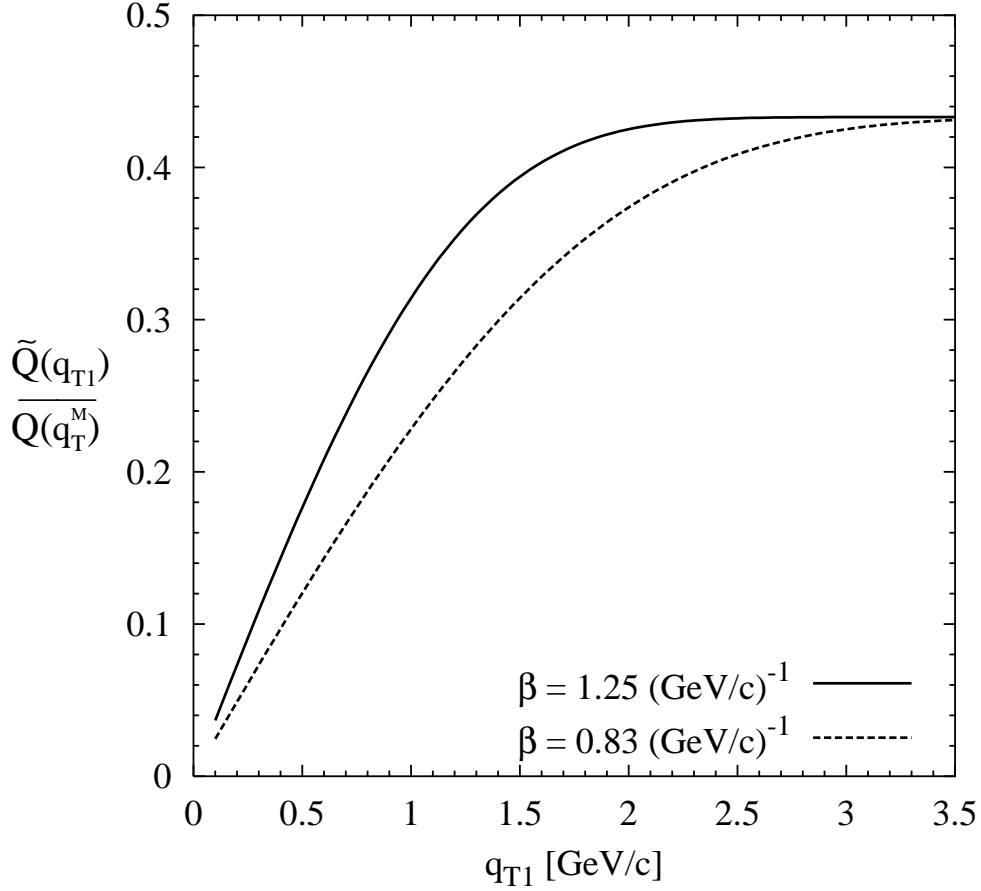,angle=-90,width=.9\textwidth}}
\caption[a7]{\small{
The factor $\tilde{\mathcal Q}(q_{T1})/{\mathcal Q}(q_T^M)$,
see Eq.s~(\ref{aveq}) and (\ref{max}), plotted as a function of $q_{T1}$,
for  $\beta=1.25$ (GeV/$c)^{-1}$ (solid line) and
$\beta=0.83$ (GeV/$c)^{-1}$ (dashed line), corresponding respectively
to $\langle\,k_\perp^2\,\rangle^{1/2}=0.8$ GeV$/c$ and
$\langle\,k_\perp^2\,\rangle^{1/2}=1.2$ GeV$/c$.
This factor can be used to obtain from the asymmetries given in Fig.s 2-5
(at fixed $\bfq_T$, $q_T=q_T^M$ and $\phi_{q_{_T}}=0$)
the corresponding asymmetries averaged over $\bfq_T$ up to
$|\bfq_T|=q_{T1}$ (see text for further details).
}}
\end{center}
\end{figure}

\end{document}